\documentclass[aps,prl,reprint,superscriptaddress,nofootinbib,longbibliography,floatfix]{revtex4-2}

\usepackage{amsmath,amssymb,bm}
\usepackage{graphicx}
\usepackage{orcidlink}
\usepackage{hyperref}
\hypersetup{hidelinks}

\newcommand{\bk}{\bm k}
\newcommand{\be}{\bm e}

\begin{document}

\title{Microscopic Compensation of Circular Optical Response in Altermagnets}

\author{Sahar Izadi Vishkayi\orcidlink{0000-0002-0314-1007}}
\email{izadi.123@gmail.com}
\affiliation{International Research Centre MagTop, Institute of Physics, Polish Academy of Sciences, Aleja Lotnik\'ow 32/46, PL-02668 Warsaw, Poland}

\author{Carmine Autieri\orcidlink{0000-0002-5008-8165}}
\email{autieri@magtop.ifpan.edu.pl}
\affiliation{International Research Centre MagTop, Institute of Physics, Polish Academy of Sciences, Aleja Lotnik\'ow 32/46, PL-02668 Warsaw, Poland}

\author{Meysam Bagheri Tagani\orcidlink{0000-0001-7065-3195}}
\email{mtagani@magtop.ifpan.edu.pl}
\affiliation{International Research Centre MagTop, Institute of Physics, Polish Academy of Sciences, Aleja Lotnik\'ow 32/46, PL-02668 Warsaw, Poland}
\affiliation{Department of Physics, University of Guilan, P. O. Box 41335-1914, Rasht, Iran}

\date{\today}

\begin{abstract}
A weak or vanishing circular optical signal need not imply weak helicity-sensitive transitions: it can be the residual of much larger opposing microscopic contributions. We resolve the finite-frequency interband Kubo response before summing over transitions and momentum, separating cancellation among transitions at fixed momentum from cancellation across the Brillouin zone. An exactly compensated $d$-wave model establishes the symmetry benchmark. In MnTe, first-principles Wannier calculations reveal a symmetry-dark N\'eel orientation with an extended sign-changing circular kernel whose bulk integral vanishes. Reorienting the N\'eel vector removes this extinction, yet the allowed response remains a small imbalance between large positive- and negative-momentum sectors; the imbalance reverses with photon energy and flips the macroscopic sign. Thus magnetic symmetry decides whether circular response is allowed, whereas microscopic compensation determines how much survives and with what sign. This distinction provides a general framework for interpreting weak, vanishing, and sign-changing magneto-optical signals in compensated magnets.
\end{abstract}

\maketitle

\textit{Introduction.---}
A small magneto-optical Hall response is often read as evidence for weak microscopic helicity sensitivity. In a multiband crystal, however, the measured coefficient is a signed sum over many optical transitions and over the Brillouin zone (BZ), so a small result may instead be the residual of much larger opposing contributions. Magnetic symmetry can determine whether the final response must vanish, but it does not determine how much microscopic weight exists before the cancellations occur.

Altermagnets provide a particularly clean setting in which to separate these questions. Their compensated collinear order produces momentum-dependent spin splitting without a net nonrelativistic magnetization~\cite{Smejkal2022PRX}, and the N\'eel-vector orientation can change the relativistic magnetic point group and hence the allowed antisymmetric optical tensor~\cite{Takahashi2025MnTe}. Circular optics in MnTe is already well established: core-level XMCD exhibits symmetry-controlled cancellation between Mn sublattices~\cite{Hariki2024XMCD}; visible-range MCD and Kerr spectra have been calculated and measured~\cite{Hubert2025OpticalMCD,Yang2026Ultrafast}; and strain-driven N\'eel rotation can continuously tune visible MCD through symmetry-forbidden orientations, with first-principles support~\cite{Liebman2026Strain}. These works establish the existence, microscopic origin, and N\'eel-orientation dependence of the macroscopic signal. They do not separate cancellation among finite-frequency valence-to-conduction transitions at fixed momentum from cancellation between inequivalent momentum regions in the subsequent Brillouin-zone sum.

Here we separate two levels of microscopic compensation: cancellation among interband transitions at fixed momentum and cancellation between momentum regions after the local channel sum. A first-principles-fitted $\mathrm{KV_2Se_2O}$ model supplies an exact symmetry-compensated benchmark~\cite{Cheng2026TB}; full-BZ spinor-Wannier calculations for MnTe then show that these two reductions can coexist with a N\'eel-controlled switch between a symmetry-forbidden and a finite macroscopic response. The distinction is independent of whether the parent altermagnetic texture is labeled $d$ or $g$ wave; those labels do not by themselves fix the relativistic optical response~\cite{Vila2025OrbitalSpin,Sivianes2026SSG}.

\begin{figure*}[t]
\centering
\includegraphics[width=\textwidth]{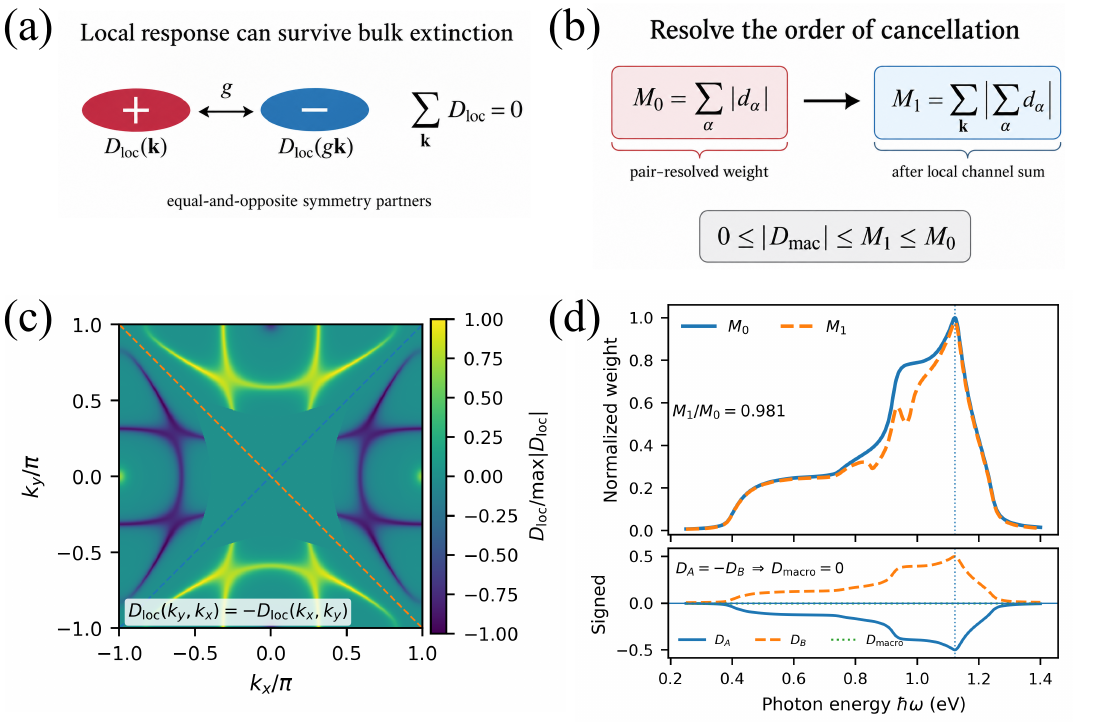}
\caption{Microscopic response and two-stage compensation. (a) A sign-changing symmetry relates equal-and-opposite local circular kernels, so a vanishing BZ sum need not imply a vanishing local response. (b) For a specified interband partition, $M_0\!\rightarrow M_1$ resolves cancellation among transitions at fixed momentum, while $M_1\!\rightarrow |D_{\rm mac}|$ resolves cancellation in the subsequent BZ sum. (c) Reduced circular-response texture $\widetilde D_{\rm loc}(k_x,k_y)$ of the $\mathrm{KV_2Se_2O}$ model at $E_*=1.12$ eV. (d) Absolute weights and signed block integrals, normalized by $\max_E M_0$. The tilde denotes a global model helicity convention; all cancellation ratios and the exact zero BZ sum are convention independent.}
\label{fig:principle}
\end{figure*}

\textit{Compensation hierarchy.---}
For equal-amplitude internal circular fields propagating along $z$, with $\be_\pm=(\hat x\pm i\hat y)/\sqrt2$ and time dependence $e^{-i\omega t}$, define~\cite{Souza2008Dichroic}
\begin{align}
D(E)\equiv D_z(E)
&=\operatorname{Re}(\be_+^\dagger\bm\sigma\be_+)
 -\operatorname{Re}(\be_-^\dagger\bm\sigma\be_-)
\notag\\
&=-2\operatorname{Im}\sigma^A_{xy}(E),
\label{eq:observable}
\end{align}
where $E=\hbar\omega$ and $\sigma^A_{xy}=(\sigma_{xy}-\sigma_{yx})/2$. Let $\alpha=(v,c)$ denote an occupied-to-empty interband transition at fixed $\bk$, and let $d_\alpha(\bk,E)$ be its signed contribution including the positive BZ quadrature weight. We define
\begin{equation}
D_{\rm loc}(\bk,E)=\sum_\alpha d_\alpha,\quad
M_0=\sum_{\bk,\alpha}|d_\alpha|,\quad
M_1=\sum_{\bk}|D_{\rm loc}|,
\label{eq:hierarchy}
\end{equation}
with $D_{\rm mac}=\sum_{\bk}D_{\rm loc}$. The triangle inequality gives
\begin{equation}
|D_{\rm mac}|\le M_1\le M_0.
\label{eq:ineq}
\end{equation}
Thus $M_1/M_0$ measures survival of the local interband sum and $|D_{\rm mac}|/M_1$ the survival of the subsequent momentum sum. Their product $U=|D_{\rm mac}|/M_0$ is a microscopic survival ratio, not an MCD percentage. For positive resonant interband weight, $M_0$ is bounded by the in-plane absorptive weight $S_\perp=\operatorname{Re}\sigma_{xx}+\operatorname{Re}\sigma_{yy}$ (End Matter), so the equal-internal-field absorption asymmetry obeys $|D_{\rm mac}|/S_\perp\le U$.

For a magnetic operation $g$ with spatial part $R$, write $g\bk=s_gR\bk$, where $s_g=+1$ for a unitary operation and $-1$ for an antiunitary one. When $R$ preserves the $z$ axis as a line,
\begin{equation}
D_{\rm loc}(g\bk,E)=\chi_gD_{\rm loc}(\bk,E),\qquad \chi_g=s_g\det(R)R_{zz}.
\label{eq:symmetry}
\end{equation}
A symmetry with $\chi_g=-1$ pairs opposite-sign local kernels and forces $D_{\rm mac}=0$; $\chi_g=+1$ removes that obstruction but does not prevent cancellation between inequivalent momentum regions.

\textit{Exact benchmark.---}
We first apply Eqs.~(\ref{eq:hierarchy})--(\ref{eq:symmetry}) to the first-principles-fitted four-band $\mathrm{KV_2Se_2O}$ Hamiltonian of Ref.~\cite{Cheng2026TB}. Its two optically closed $2\times2$ blocks obey $\widetilde D_{\rm loc}(k_x,k_y,E)=-\widetilde D_{\rm loc}(k_y,k_x,E)$, which forces the BZ integral to vanish at every sampled photon energy [Fig.~\ref{fig:principle}(c)]. At $E_*=1.122$ eV, $M_1/M_0=0.98077$: little weight is lost in the local channel sum, whereas momentum integration cancels the remainder exactly [Fig.~\ref{fig:principle}(d)]. Both SOC amplitudes of the model are required for a nonzero circular vertex. The analytic two-band identity, sign convention, and convergence from $201^2$ to $801^2$ meshes are given in the Supplemental Material (SM)~\cite{SM}.

\begin{figure}[b]
\centering
\includegraphics[width=\columnwidth]{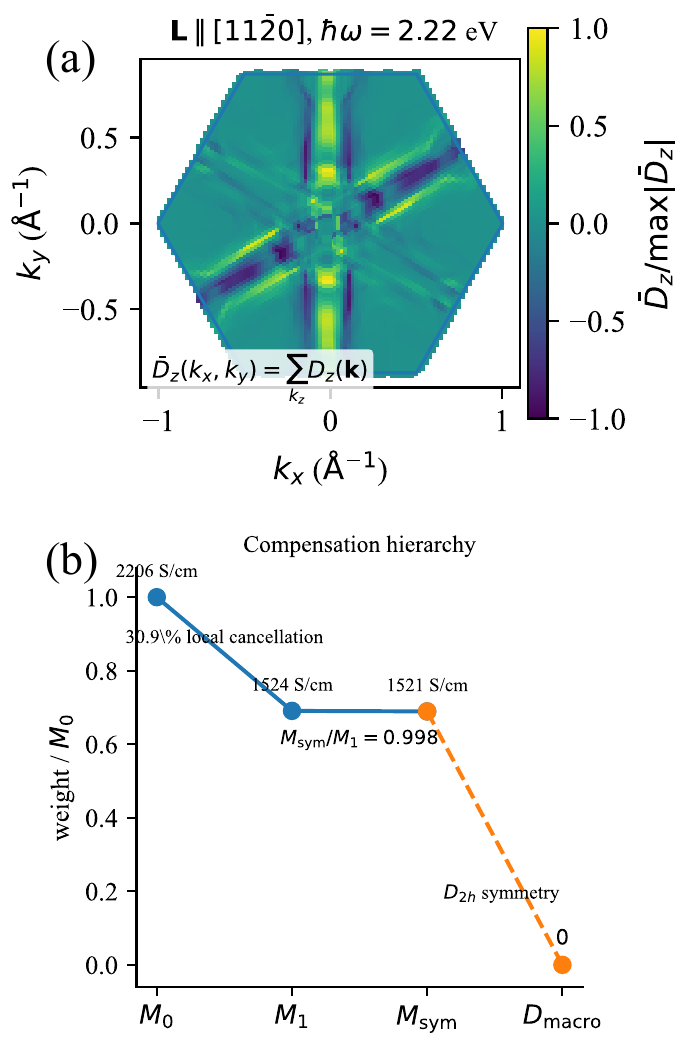}
\caption{Symmetry extinction hides a finite microscopic circular kernel in MnTe. (a) Raw $k_z$-summed kernel for $\mathbf L\parallel[11\bar20]$ at $2.22$ eV on the $72\times72\times48$ mesh. The sign-changing $C_{2n}$ partner relation is satisfied to a normalized $L_1$ mismatch of $1.91\%$ (SM). (b) The physical compensation hierarchy contains only the local interband reduction $M_0\to M_1$ followed by the symmetry-forced macroscopic zero; symmetry projection is used separately as a numerical diagnostic in the SM. Values above the points are in S/cm.}
\label{fig:dark}
\end{figure}

\textit{Symmetry-dark MnTe.---}
For bulk $\alpha$-MnTe we construct a 36-function spinor-Wannier representation from PBE+$U$+SOC calculations and evaluate the full-BZ interband conductivity with the native Wannier90 Berry connection~\cite{Souza2001Disentanglement,Yates2007Interpolation,Mostofi2014Wannier}. The two N\'eel orientations are generated from independent relativistic calculations rather than by rotating a postprocessing axis. Electronic-structure, Wannier, optical-closure, and mesh checks are documented in the SM~\cite{SM}.

MnTe has a three-dimensional nonrelativistic $g$-wave parent texture, while SOC and the N\'eel orientation determine the exact relativistic magnetic point group~\cite{Takahashi2025MnTe,Fakhredine26}. For the ideal $\mathbf L\parallel[11\bar20]$ orientation the relevant group is all-unitary $D_{2h}$. An in-plane twofold rotation $C_{2n}$ has $\chi_g=-1$ and therefore enforces $D_{\rm mac}=0$. Figure~\ref{fig:dark}(a) shows that this macroscopic extinction coexists with a pronounced sign-changing microscopic kernel. At $2.22$ eV, $M_0=2206.34$ S/cm and $M_1=1523.97$ S/cm, so $30.93\%$ of the specified transition-resolved absolute weight cancels already at fixed momentum before the remaining kernel is extinguished in the BZ sum. The ideal zero is fixed by symmetry; projection onto the ideal character and the raw partner tests are used only to quantify numerical symmetry fidelity (SM)~\cite{SM}.

\begin{figure*}[t]
\centering
\includegraphics[width=\textwidth]{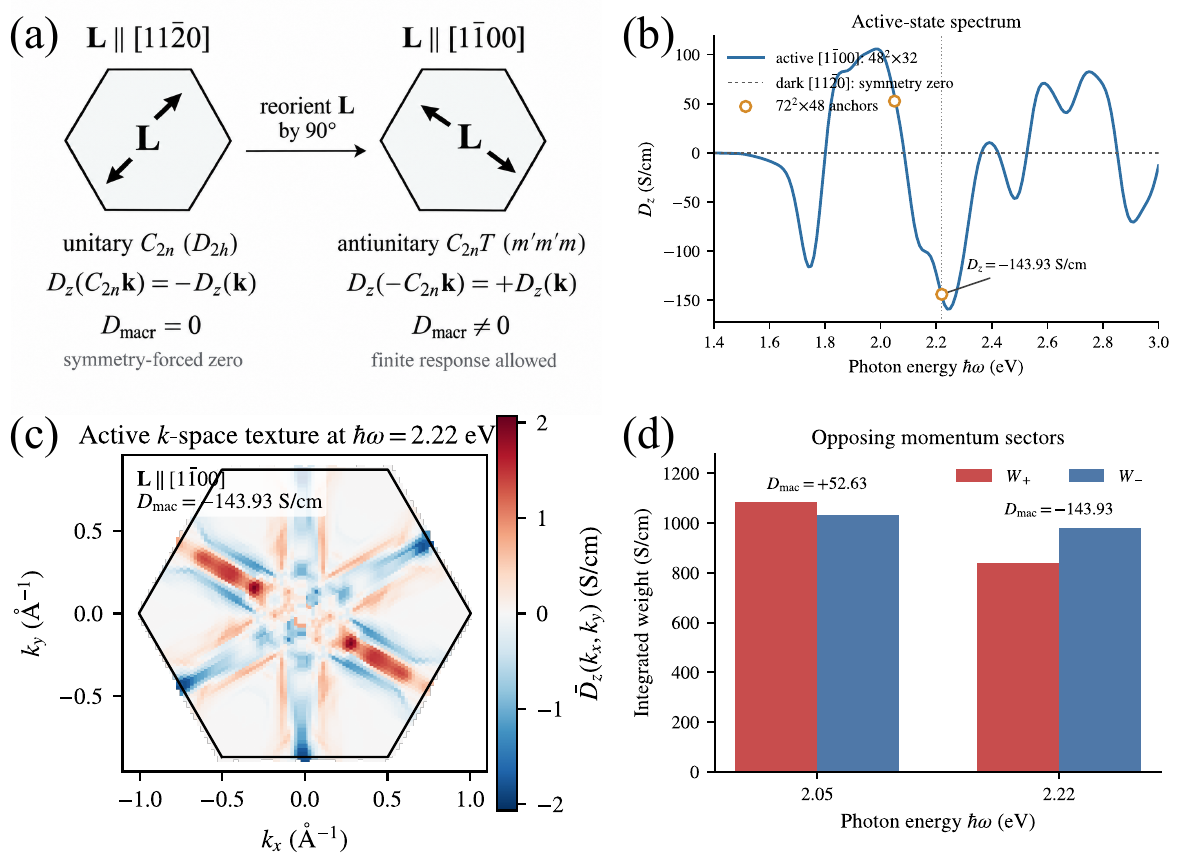}
\caption{N\'eel reorientation removes exact extinction but not microscopic compensation. (a) For $\mathbf L\parallel[11\bar20]$ the relevant twofold operation is unitary and sign changing, whereas for $\mathbf L\parallel[1\bar100]$ it is antiunitary and sign preserving. (b) Active-state spectrum on the $48^2\times32$ mesh with independent $72^2\times48$ anchors; the dark line is the ideal symmetry result. (c) Raw active $k_z$-summed kernel at $2.22$ eV. (d) Dense-mesh positive and negative momentum-sector weights $W_+$ and $W_-$. Their small imbalance reverses between 2.05 and 2.22 eV, reversing the sign of $D_{\rm mac}=W_+-W_-$.}
\label{fig:active}
\end{figure*}

\textit{N\'eel reorientation and residual compensation.---}
For $\mathbf L\parallel[1\bar100]$ the relativistic magnetic point group becomes $m'm'm$~\cite{Takahashi2025MnTe}. The relevant in-plane twofold operation is now antiunitary, $C_{2n}{\cal T}$, with $\chi_g=+1$ [Fig.~\ref{fig:active}(a)]. Symmetry therefore permits a finite axial-$z$ circular response. The independently generated active-state calculation gives $D_z=-143.93$ S/cm at $2.22$ eV on the $72\times72\times48$ mesh, within $1.42\%$ of the $48\times48\times32$ value [Fig.~\ref{fig:active}(b)]. At $2.05$ eV the dense-mesh response has the opposite sign, and the sampled spectrum crosses zero between $2.08$ and $2.09$ eV.

The finite active response remains a small residual of a much larger microscopic kernel. At $2.22$ eV, $M_0=2518.62$ S/cm and $M_1=1819.91$ S/cm: $27.74\%$ cancels in the local interband sum and only $|D_{\rm mac}|/M_1=7.91\%$ of the post-channel-sum weight survives momentum integration. To expose the origin of the spectral sign, define the total positive and negative momentum weights
\begin{equation}
W_+=\sum_{\bk}\max(D_{\rm loc},0),\qquad
W_-=-\sum_{\bk}\min(D_{\rm loc},0),
\label{eq:wpm}
\end{equation}
so that $M_1=W_++W_-$ and $D_{\rm mac}=W_+-W_-$. At the two dense anchors the ordering of $W_+$ and $W_-$ reverses [Fig.~\ref{fig:active}(d)]: the sign change therefore reflects a spectral redistribution between two large opposing momentum-space sectors rather than the disappearance of one sector. N\'eel reorientation removes exact symmetry extinction, but it does not ``release'' the full microscopic circular weight. Additional $k_z$- and channel-resolved analyses are given in the SM~\cite{SM}.

\textit{Implications.---}
The compensation hierarchy clarifies what a weak magneto-optical coefficient does and does not imply. Earlier MnTe work established site-resolved cancellation in core-level XMCD~\cite{Hariki2024XMCD}, the microscopic origin of visible-range MCD/Kerr~\cite{Hubert2025OpticalMCD}, and the strong N\'eel-orientation dependence of visible and ultrafast magneto-optical signals~\cite{Yang2026Ultrafast,Liebman2026Strain}. The present decomposition resolves a different microscopic level of the valence-to-conduction charge response: first the signed competition among interband transitions at each momentum, and then the cancellation between momentum regions. Symmetry decides whether the second stage is exact; matrix elements, occupations, and resonance energies determine how close to cancellation an allowed response remains.

For equal internal circular-field amplitudes, the absorbed-power asymmetry is $D_{\rm mac}/S_\perp$ rather than $U$; external MCD, Kerr, or Faraday signals additionally depend on the full dielectric tensor and sample electrodynamics. The quoted cancellation fractions are therefore microscopic diagnostics of the stated independent-particle transition partition and broadening, not universal material constants. Within that scope, the same hierarchy distinguishes weak microscopic helicity sensitivity from strong microscopic response hidden by compensation, providing a direct way to interpret small and sign-changing optical Hall signals in compensated magnets.

\begin{acknowledgments}
M.B.T. acknowledges support from the Polish National Agency for Academic Exchange (NAWA) under the ULAM program, project BPN/ULM/2025/1/00156/U/00001, and from the Iran National Science Foundation under project 4043973. This work was supported by the Foundation for Polish Science project ``MagTop,'' FENG.02.01-IP.05-0028/23, co-financed by the European Union under the European Funds for a Smart Economy Program 2021--2027. We acknowledge computing resources of the Interdisciplinary Centre for Mathematical and Computational Modelling, University of Warsaw (grants g91-1418, g91-1419, g96-1808, g96-1809, and g103-2540), and the Poznan Supercomputing and Networking Center (grants pl0267-01, pl0365-01, pl0807, and pl0471-01).
\end{acknowledgments}

\noindent\textit{Data availability.---} The numerical data and analysis scripts needed to reproduce the reported figures and compensation diagnostics will be made available with the published article; during review they are available from the corresponding authors upon reasonable request.

\nocite{Amin2024MnTe,BagheriTagani2026RankSelective,GonzalezBetancourt2023AHE,Jiang2025KV2Se2O,Lee2024MnTe}
\bibliography{references_prl_final}

\section*{End Matter}

\textit{Kubo kernel and optical convention.---}
For an occupied band $v$ and an empty band $c$ at momentum $\bk$, let $\epsilon_v(\bk)$ and $\epsilon_c(\bk)$ be their eigenenergies, $f_v$ and $f_c$ their occupation factors, and $r^i_{vc}(\bk)=i\langle u_{v\bk}|\partial_{k_i}u_{c\bk}\rangle$ the interband Berry connection. With $\Delta_{cv}=\epsilon_c-\epsilon_v>0$ and $\sum_{\bk}w_{\bk}=1$, the resonant contribution used in the decomposition is~\cite{Souza2008Dichroic,Yates2007Interpolation}
\begin{equation}
 d_{vc}(\bk,E)
 =-\frac{2\pi e^2}{\hbar V_c}
 w_{\bk}(f_v-f_c)\Delta_{cv}
 \operatorname{Im}[r^x_{vc}r^y_{cv}]\,
 \delta_\eta(E-\Delta_{cv}),
\label{eq:kubo}
\end{equation}
where $V_c$ is the unit-cell volume and $e>0$. The BZ quadrature weight is included once in $d_{vc}$ and is not reapplied in later sums. Reversing the helicity convention reverses every signed kernel and $D_{\rm mac}$ but leaves $M_0$, $M_1$, and $U$ unchanged.

For positive resonant weight,
\[
2|\operatorname{Im}(r^x_{vc}r^y_{cv})|\le |r^x_{vc}|^2+|r^y_{vc}|^2,
\]
which gives $M_0\le S_\perp$ for the same transition set and broadening, with $S_\perp=\operatorname{Re}\sigma_{xx}+\operatorname{Re}\sigma_{yy}$ and $\sigma_{xx},\sigma_{yy}$ the diagonal optical-conductivity components. If the two internal circular fields have the same amplitude $|E_{\rm int}|$, their cycle-averaged absorbed powers are
\[
p_\pm=\frac{|E_{\rm int}|^2}{2}\operatorname{Re}(\be_\pm^\dagger\bm\sigma\be_\pm),
\]
and therefore
\begin{equation}
\frac{p_+-p_-}{p_++p_-}
=\frac{D_{\rm mac}}{S_\perp}
=\operatorname{sgn}(D_{\rm mac})
\frac{M_0}{S_\perp}\frac{M_1}{M_0}\frac{|D_{\rm mac}|}{M_1}.
\label{eq:experimental}
\end{equation}
This factorization separates microscopic helicity-sensitive weight from its survival under the two signed sums. External reflection, transmission, Kerr, and Faraday observables require electrodynamic propagation through the sample.

\textit{Channel resolution and symmetry diagnostics.---}
For nondegenerate states, $r^x_{vc}r^y_{cv}$ is invariant under independent Bloch-state phase choices. At an exact degeneracy, individual state-to-state absolute weights depend on the basis chosen within the degenerate subspace; only the complete multiplet tensor, e.g. $\operatorname{Tr}[P_Vr_xP_Cr_yP_V]$, is invariant. Hence $M_0$, the local-cancellation fraction, and $U$ are diagnostics of the explicitly stated transition partition, whereas the complete $D_{\rm loc}$, $M_1$, and $D_{\rm mac}$ are unchanged by relabeling inside a fully retained multiplet. Detailed multiplet and energy-rank conventions are given in the SM~\cite{SM}.

On a uniform mesh closed under the magnetic-group operations, the ideal symmetry projector is
\begin{equation}
(\mathcal PD_{\rm loc})(\bk)=\frac{1}{|G|}\sum_{g\in G}\chi_gD_{\rm loc}(g\bk).
\label{eq:projection}
\end{equation}
We use this only as a diagnostic of the ideal magnetic character. Numerical symmetry fidelity is assessed independently by the partner mismatch, sign-adjusted correlation, and $\|D_{\rm loc}-\mathcal PD_{\rm loc}\|_1/M_1$; complete values are reported in the SM~\cite{SM}. Projection is not used as evidence for convergence of the electronic structure or optical matrix elements.

\textit{First-principles and Wannier implementation.---}
Bulk $\alpha$-MnTe is treated in the NiAs structure using VASP with PAW potentials, PBE exchange correlation, noncollinear SOC, and Liechtenstein DFT+$U$ with $U=4.00$ eV and $J=0.90$ eV, together with a 600-eV plane-wave cutoff~\cite{Blochl1994PAW,Kresse1996VASP,PBE1996,Liechtenstein1995U}. The two N\'eel orientations are independent relativistic electronic calculations with spatial symmetry disabled. A $12\times12\times8$ source mesh is disentangled from 40 spinor bands to 36 Wannier functions of Mn-$s,d$ and Te-$p$ character~\cite{Marzari1997Wannier,Souza2001Disentanglement,Mostofi2014Wannier}. The resulting representation contains 22 occupied and 14 empty states.

The optical response is evaluated with a modified Wannier90 2.1 \texttt{postw90} implementation using the native Berry connection constructed from Bloch overlaps. The added channel- and momentum-resolved accumulators do not alter the optical matrix elements and reproduce the native Kubo response to numerical precision. Production spectra use the normalized Gaussian $\delta_\eta(x)=\exp[-(x/\eta)^2]/(\sqrt\pi\eta)$ with $\eta=50$ meV. This width is a numerical spectral regularization, not a measured lifetime; consequently, precise values of the transition-resolved cancellation fraction should be understood at the stated resolution. Dense-mesh comparisons hold the Hamiltonian, occupations, optical manifold, and broadening fixed. The SM reports source-mesh, interpolation, direct-gap, symmetry, and optical-closure tests~\cite{SM}.

The active calculation carries a small residual net spin moment (SM). SOC-induced weak ferromagnetism and finite orbital magnetization have been discussed independently for MnTe~\cite{Ye2026OrbitalMagnetization}, and magnetic-moment and N\'eel-order contributions can share the same circular-optical symmetry in this geometry. We therefore do not infer the origin of the residual moment from smallness alone and do not decompose the calculated optical response into separate weak-ferromagnetic, orbital-magnetization, and compensated-order pieces. Existing visible-range calculations constrained to zero net magnetization nevertheless reproduce the N\'eel-dependent MCD phenomenology~\cite{Liebman2026Strain}.

\textit{Model implementation.---}
The $\mathrm{KV_2Se_2O}$ benchmark uses the SOC-refitted parameter set and four-orbital basis of Ref.~\cite{Cheng2026TB}, $\mu=0$, zero-temperature occupations, and Lorentzian half-width $\gamma=0.015$ eV. Its two optically closed blocks comprise orbitals $(0,3)$ and $(1,2)$. For each block we evaluate
\[
P^{cv}_{xy}=\langle c|\partial_{k_x}h|v\rangle\langle v|\partial_{k_y}h|c\rangle,
\]
and use the reduced occupied-to-empty kernel
\[
\widetilde d_b(\bk,E)=-\frac{2\operatorname{Im}P^{cv}_{xy}}{\Delta_b}L_\gamma(E-\Delta_b),\qquad
L_\gamma(x)=\frac{\gamma/\pi}{x^2+\gamma^2}.
\]
A common positive dimensional prefactor is omitted. The matrix-element ordering in this archived model convention is the complex conjugate of Eq.~(\ref{eq:kubo}), so its signed kernel carries an overall opposite helicity sign; all absolute weights, compensation ratios, and exact partner antisymmetries are unaffected. The complete Hamiltonian, parameters, and analytic/eigenvector vertex checks are given in the SM~\cite{SM}.

\end{document}